\documentclass[conference]{IEEEtran}
\IEEEoverridecommandlockouts

\usepackage[mode=buildnew]{standalone}
\usepackage{amssymb}
\usepackage{cite}
\usepackage{amsmath,amssymb,amsfonts}
\usepackage{algorithmic}
\usepackage{algorithm}
\usepackage{graphicx}
\usepackage{textcomp}
\usepackage{xcolor}
\usepackage{caption}

\def\BibTeX{{\rm B\kern-.05em{\sc i\kern-.025em b}\kern-.08em
    T\kern-.1667em\lower.7ex\hbox{E}\kern-.125emX}}  
\begin{document}

\title{Deep Unfolded Fractional Optimization for Maximizing Robust Throughput in 6G Networks}

\author{\IEEEauthorblockN{Anh Thi Bui\IEEEauthorrefmark{1}, Robert-Jeron Reifert\IEEEauthorrefmark{1}, Hayssam Dahrouj\IEEEauthorrefmark{2} and Aydin Sezgin\IEEEauthorrefmark{1}}
	\IEEEauthorblockA{\IEEEauthorrefmark{1}Ruhr University Bochum, Germany, \hspace*{.2cm}\IEEEauthorrefmark{2}University of Sharjah, United Arab Emirates.}
	\thanks{
		This work was supported in part by the German Federal Ministry of Education and Research (BMBF) in the course of the 6GEM Research Hub under grant 16KISK037.}
}

\maketitle

\begin{abstract}
The sixth-generation (6G) of wireless communication networks aims to leverage artificial intelligence tools for efficient and robust network optimization. This is especially the case since traditional optimization methods often face high computational complexity, motivating the use of deep learning (DL)-based optimization frameworks. In this context, this paper considers a multi-antenna base station (BS) serving multiple users simultaneously through transmit beamforming in downlink mode. To account for robustness, this work proposes an uncertainty-injected deep unfolded fractional programming (UI-DUFP) framework for weighted sum rate (WSR) maximization under imperfect channel conditions. The proposed method unfolds fractional programming (FP) iterations into trainable neural network layers refined by projected gradient descent (PGD) steps, while robustness is introduced by injecting sampled channel uncertainties during training and optimizing a quantile-based objective. Simulation results show that the proposed UI-DUFP achieves higher WSR and improved robustness compared to classical weighted minimum mean square error, FP, and DL baselines, while maintaining low inference time and good scalability. These findings highlight the potential of deep unfolding combined with uncertainty-aware training as a powerful approach for robust optimization in 6G networks.
\end{abstract}


\section{Introduction}
The rapid evolution of wireless communication systems continues to shape the digital foundation of modern society. While the fifth-generation (5G) of wireless networks has enabled enhanced mobile broadband and massive connectivity, the envisioned sixth-generation (6G) of wireless systems aims to achieve ubiquitous intelligence and robustness by tightly integrating artificial intelligence (AI) with data-driven optimization methods \cite{ahokangas2024changing}. Future 6G networks are expected to operate under highly dynamic and uncertain environments, where robust, efficient, and reliable resource allocation remains a fundamental challenge. To address this challenge, this work presents a deep unfolding-based optimization framework that integrates uncertainty injection to enable robust transmit beamforming and maximize the weighted sum rate (WSR). Specifically, we consider a multi-antenna base station (BS) that simultaneously serves multiple users in the downlink under imperfect channel conditions. 

While the literature of wireless networks design is rich with optimization algorithms, fractional programming (FP) recently proves to be quite useful in maximizing ratio-utilities such as energy or spectral efficiency, often formulated as WSR maximization problems \cite{shen2018fractional}. Classical FP methods, including Dinkelbach algorithm, are efficient for single-ratio problems but become computationally intensive when extended to multi-ratio formulations. The quadratic transform (QT) introduced in \cite{shen2018fractional} reformulates multi-ratio problems into tractable iterative convex subproblems, finding broad applications in beamforming and energy-efficient communications \cite{9813735,8931772}.

Despite their provable numerical benefits, traditional FP algorithms remain highly iterative, limiting their applicability in real-time or rapidly varying 6G scenarios. To address such a limitation, deep unfolded FP (DUFP) frameworks have emerged recently, where each iteration of FP is unfolded into a trainable neural network layer. 
This approach leverages deep learning (DL) to enable near-optimal solutions for complex, non-convex problems and large-scale processing tasks in wireless networks \cite{qin2019deep,daiDLwireless,dahrouj2021overview}.
The DUFP framework combines the interpretability and convergence guarantees of FP optimization with the inference speed of DL. However, while DUFP frameworks, e.g., \cite{zhao2023deep,9524496}, show promising performance under ideal network conditions, their extension to imperfect channel state information remains largely unexplored.

In fact, robustness is increasingly recognized as a key design objective for AI-native 6G networks, which are expected to operate in highly dynamic, heterogeneous, and mission-critical environments \cite{IMT-2030}. The combination of diverse service requirements, mobility patterns, and decentralized intelligence then introduces large uncertainty into the network processes. In practical deployments, therefore, it becomes indispensable to account for the uncertainty of wireless channels, interference, and even computational workloads \cite{yang2008distributed,cui2023uncertainty,kang2023deep,reifert2024robust}. Learning-based models have been widely adopted to address uncertainty, with reliability further improved through robust optimization, noise injection, and adversarial training. However, the integration of learning-based models within deep unfolded optimization frameworks remains largely unexplored \cite{deka2025comprehensive}.\looseness-1

Despite the recent progress in using deep unfolding in wireless networks optimization, efforts to incorporate robustness into the available optimization frameworks remain scarce \cite{wang2025lightweight,10013710,10540175}. In \cite{wang2025lightweight}, a reinforcement-learning-aided weighted minimum mean square error (WMMSE) precoding scheme is unfolded for robustness under channel errors. The work in \cite{10013710} considers deep unfolding without FP, focusing on the average (i.e., not worst-case), performance. Meanwhile, \cite{10540175} uses an unfolded architecture and focuses on sensing-related robustness metrics. Generally speaking, the above studies highlight the promises of using robust deep unfolding but do not investigate the numerical prospects of robust FP-based optimization.

Unlike the above references, this paper proposes an uncertainty-injected deep unfolded fractional programming (UI-DUFP) framework for maximizing the robust WSR in 6G networks. The proposed approach unfolds the FP iterations into differentiable layers optimized via projected gradient descent (PGD). Robustness is then introduced by injecting channel uncertainties during training and optimizing model parameters using a quantile-based objective that focuses on worst-case performance. Simulation results show that the proposed framework achieves higher robust WSR and lower inference time compared to classical WMMSE and FP algorithms, thereby demonstrating its potential in providing a robust framework for optimizing real-time, uncertainty-aware 6G networks.

\section{System Model and Problem Formulation}
\begin{figure}[t]
\centering
\includegraphics[width=1\linewidth]{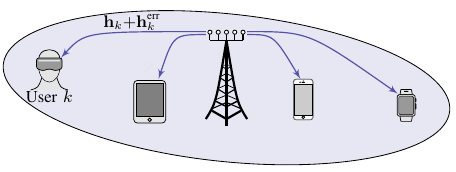}\vspace*{-.2cm}
\caption{Network setup with a multi-antenna BS serving 4 users, illustrating the imperfect channel knowledge.}
\label{model}
\vspace*{-.4cm}
\end{figure}

In this paper, we consider a single BS equipped with $L$ transmit antennas that serves $K$ single-antenna users in the downlink. The set of users is denoted by $\mathcal{K}=\{1,\ldots,K\}$. Each user receives an independent data stream transmitted via linear precoding with spatial multiplexing. The channel coefficient between the BS's antenna $l$ and user $k$ is denoted by $h_{l,k}$. Fig.~\ref{model} illustrates an example setup with four users.
The received baseband signal at user $k$ is expressed as
\begin{equation}\label{eq:1}
y_k = \sideset{}{_{j\in\mathcal{K}}}\sum \mathbf{h}_k^H\mathbf{v}_j d_j + n_k,
\end{equation}
where $\mathbf{h}_k = [h_{1,k},\dots,h_{L,k}]^T \in \mathbb{C}^{L \times 1}$ denotes the aggregated channel vector between the BS and user $k$, $\mathbf{v}_j \in \mathbb{C}^{L \times 1}$ is the beamforming vector for user $j$, $d_k \sim \mathcal{CN}(0,1)$ represents the transmitted data symbols, and $n_k \sim \mathcal{CN}(0, \sigma^2)$ is the additive white Gaussian noise. The full channel matrix is denoted by $\mathbf{H} = [\mathbf{h}_1, \mathbf{h}_2, \dots, \mathbf{h}_K] \in \mathbb{C}^{L \times K}$.

The instantaneous signal-to-interference-plus-noise ratio (SINR) of user $k$ is given by
\begin{equation}\label{eq:2}
{\Gamma}_k = \frac{|\mathbf{h}_k^H\mathbf{v}_k|^2}{ \sum_{j\in\mathcal{K}\backslash\{k\}}|\mathbf{h}_k^H\mathbf{v}_j|^2+\sigma^2}.
\end{equation} 
and the corresponding spectral efficiency is
\begin{equation}\label{eq:3}
    r_k =\log_2(1 + \Gamma_k).
\end{equation}

In practice, the BS has only imperfect knowledge of the true channel coefficients, which are subject to estimation errors and environmental dynamics \cite{cui2023uncertainty}. The estimated channel vector $\mathbf{\hat{h}}_k$ can be modeled as\vspace*{-.1cm}
\begin{align}\label{eq:4}
\mathbf{\hat{h}}_k 
= {\mathbf{h}}_k + \mathbf{h}_k^{err},\\[-.65cm]\nonumber
\end{align}
where the error term $\mathbf{h}_k^{err} \sim \mathcal{CN}(0, \sigma_h^2\mathbf{I}_L)$ captures independent estimation uncertainties across antennas \cite{cui2023uncertainty}, and $\mathbf{I}_L$ is the $L\times L$ identity matrix. The corresponding estimated channel matrix is $\mathbf{\hat{H}}\ = [{\mathbf{\hat{h}}_1},{\mathbf{\hat{h}}}_2, \dots, \mathbf{\hat{h}}_K] $.

The uncertainty model depicted in \eqref{eq:4} serves as the foundation for the robustness analysis and the proposed uncertainty-injected deep unfolded fractional programming framework in the subsequent sections of the paper.

\subsection{Problem Formulation}
The problem considered in this paper can be formulated as the following classical WSR maximization problem:\vspace*{-.1cm}
\begin{subequations}\label{eq:5}
\begingroup
\addtolength{\jot}{-.1cm}
\begin{align}
	\underset{\mathbf{V}}{\text{max}}\quad &\sideset{}{_{k\in\mathcal{K}}}\sum w_k r_k  \tag{\ref{eq:5}} \\
    \text{s.t.} \quad\, & \mathrm{Tr}(\mathbf{V} \mathbf{V}^H) \leq P_{\max}, \label{eq:Pmax}\\[-.55cm]\nonumber
\end{align}
\endgroup
\end{subequations}
where $\mathbf{V} = [\mathbf{v}_1, \mathbf{v}_2, \dots, \mathbf{v}_K]$ contains the beamforming vectors, $w_k$ denotes the user priority weight, and $P_{\max}$ represents the total transmit-power budget. The trace term $\mathrm{Tr}(\mathbf{V} \mathbf{V}^H)$ corresponds to the total radiated power.

The problem in \eqref{eq:5} is non-convex and NP-hard due to the coupled interference terms in the SINR expression \eqref{eq:2}, as well as the imperfect channel knowledge \cite{liu2010coordinated}. To obtain a tractable reformulation of problem \eqref{eq:5}, one can first use the QT to handle its intricacy \cite{shen2018fractional}. To this end, let $\text{S}_k=|\mathbf{h}_k^H\mathbf{v}_k|^2$ and $\text{I}_k = \sum_{j\in\mathcal{K}\backslash\{k\}}^{K}|\mathbf{h}_k^H\mathbf{v}_j|^2+\sigma^2$. The transformed equivalent of the objective in \eqref{eq:5} becomes\vspace*{-.1cm}
\begin{align}
R_Q(\mathbf{V}, \mathbf{g}, \mathbf{u}) = \sum_{k\in\mathcal{K}} \Big(2 u_k \sqrt{w_k (1+g_k) S_k} - u_k^2 (S_k + I_k) \Big), \nonumber\\[-.4cm]\label{eq:6}\\[-.7cm]\nonumber
\end{align}
where $u_k$ and $g_k$ are auxiliary variables introduced to decouple the fractional terms. Based on \eqref{eq:6}, the FP-based downlink beamforming problem can be written as\vspace*{-.1cm}
\begin{subequations}\label{eq:7}
\begingroup
\addtolength{\jot}{-.1cm}
\begin{align}
	\underset{\mathbf{V},\mathbf{g}, \mathbf{u}}{\text{max}}\quad & R_Q(\mathbf{V}, \mathbf{g}, \mathbf{u})  \tag{\ref{eq:7}} \\
    \text{s.t.} \quad\; & \eqref{eq:Pmax}, \nonumber \\
    & g_k \in \mathbb{R}, \; u_k \in \mathbb{C}, &&\forall k\in\mathcal{K},\label{eq:gamma_u}\\[-.6cm]\nonumber
\end{align}
\endgroup
\end{subequations} 
where the set of optimization variables includes the beamforming vectors $\mathbf{V}$, in addition to the auxiliary variables $\mathbf{g} = [g_1, \dots, g_K]^T$ and $\mathbf{u} = [u_1, \dots, u_K]^T$. Although the QT enables efficient iterative updates, the resulting solution does not account for the channel uncertainty as per \eqref{eq:4}.

To incorporate robustness, we adopt a probabilistic quantile-based criterion \cite{cui2023uncertainty}, where the achievable WSR becomes a random variable due to uncertain channels. Let $R_{\max}^\gamma$ denote the $\gamma$-th quantile of the achievable ${R}_Q(\mathbf{V}, \mathbf{g}, \mathbf{u})$, conditioned on the estimated channels $\mathbf{\hat{h}}_k$. The robustness constraint can then be written as: \vspace*{-.1cm}
\begin{align}\label{eq:8}
    \mathbf{Pr}[{R}_Q(\mathbf{V}, \mathbf{g}, \mathbf{u}) < R_{\max}^\gamma | {\mathbf{h}}_k, \forall k \in \mathcal{K}] \leq \gamma,\\[-.6cm]\nonumber
\end{align}
which guarantees that the achievable WSR would fall below  $R_{\max}^\gamma$ with probability at most $\gamma$. For the remaining $1-\gamma$ fraction of realizations, the achievable WSR meets or exceeds the $\gamma$. The resulting robust WSR maximization problem is formulated as
\begin{subequations}\label{eq:9}
\begingroup
\addtolength{\jot}{-.1cm}
\begin{align}
	\underset{\mathbf{V},\mathbf{g},\mathbf{u}}{\text{max}}\quad 
    &R_{\max}^\gamma
    \tag{\ref{eq:9}} \\
    \text{s.t.} \quad\; & \eqref{eq:4}, \eqref{eq:Pmax}, \eqref{eq:gamma_u}, \eqref{eq:8}.\nonumber 
\end{align}
\endgroup
\end{subequations}
The paper, therefore, next seeks to maximize the robust WSR of the network under channel uncertainty by developing an UI-DUFP framework that efficiently solves the complex optimization problem in \eqref{eq:9}.

\section{Robust Deep Unfolding}
To solve the robust optimization problem in \eqref{eq:9}, we first derive the mathematical update equations required to obtain the optimal auxiliary variables and beamforming vectors. We then describe how these iterative updates are unfolded into the DUFP framework \cite{zhao2023deep}. Specifically, the variables $\mathbf{V}$, $\mathbf{g}$, and $\mathbf{u}$ are updated sequentially in an iterative manner as described next. 
\subsection{DUFP Framework} 
Firstly, the auxiliary variables $u_k$, $\forall k\in\mathcal{K}$, are updated according to
\begin{equation}\label{eq:10}
    u^*_k = \frac{\sqrt{w_k(1+g_k)S_k}}{S_k + I_k}, \quad \forall k\in\mathcal{K}.
\end{equation}
Next, $g_k$, $\forall k\in\mathcal{K}$, are obtained by setting
\begin{equation}\label{eq:11}
    g^*_k = \frac{S_k}{I_k}.
\end{equation}
Afterward, the beamforming vectors $\mathbf{v}_k(\nu)$ become functions of the Lagrange multiplier $\nu \geq 0$, which is associated with the total power constraint \eqref{eq:Pmax} and determined from
\begin{equation}\label{eq:12}
    \nu^* =\min \left\{ \nu \geq 0: \sum_{k=1}^{K} \| \mathbf{v}_k(\nu) \|_2^2 \;\leq\; P_{\max} \right\}.
\end{equation} 
The beamforming vectors are then obtained as the optimal solution of a convex subproblem with respect to $\mathbf{V}$ for fixed $\mathbf{g}$ and $\mathbf{u}$, yielding
\begin{equation}\label{eq:13}
    \mathbf{v}_k(\nu) = {\sqrt{w_k(1+g_k)}u_k}\Bigg({\sum_{j\in\mathcal{K}} |u_j|^2 \, \mathbf{h}_j \mathbf{h}_j^{H} + \nu \mathbf{I}_L}\Bigg)^{-1}\mathbf{h}_k.
\end{equation}
It is obvious from \eqref{eq:13} that computing the beamforming matrix $\mathbf{V}$ requires iterative updates involving matrix inversions, eigendecompositions, and binary search operations, resulting in high computational complexity. This motivates the use of deep unfolding to approximate these updates through trainable neural network layers. Let the feasible set corresponding to the power constraint be defined as
\begin{equation}\label{eq:14}
    \mathcal{C} = \{ \mathbf{V} \mid \mathrm{Tr}(\mathbf{V} \mathbf{V}^H) \le P_{\max} \}. 
\end{equation}  
As proposed in \cite{zhao2023deep}, the DUFP framework reduces the computational burden of iterative FP algorithms by representing $M$ iterations as a sequence of trainable layers. Within each layer, $N$ PGD steps are executed to refine the beamforming updates while satisfying the transmit power constraint.

Specifically, the beamforming vector for user $k$ at the $n$-th PGD step of the $m$-th layer is denoted by $\mathbf{v}_{m,k}^{(n)}$, and the corresponding beamforming matrix is $\mathbf{V}_{m}^{(n)} = [\mathbf{v}_{m,1}^{(n)},\dots,\mathbf{v}_{m,K}^{(n)}]$ for all users. The PGD update at step $n$ of layer $m$ is formulated as\looseness-1
\begin{align}\label{eq:15}
\begin{split}
    & \mathbf{V}_m^{(n)} = \mathbf{V}_m^{(n-1)} - \mu_m^{(n)}\nabla_\mathbf{V}\! \ R_Q(\mathbf{V}_m^{(n-1)}, \mathbf{g}, \mathbf{u}) ,\\
   & \mathbf{V}_m^{(n)} = \Omega_C\{\mathbf{V}_m^{(n)}\},   
\end{split}
\end{align}
where $\mu_m^{(n)}$ denotes the step size of the PGD update at layer $m$, and $\nabla_\mathbf{V} R_Q(\mathbf{V}_m^{(n)}, \mathbf{g}, \mathbf{u})$ represents the gradient of the objective function with respect to the beamforming matrix. The PGD iterations are followed by a projection step $\Omega_C\{\mathbf{V}_m^{(n)}\}$ to satisfy the transmit power constraint as follows
\begin{align}
\Omega_C\big\{\mathbf{V}_m^{(n)}\big\} =
\begin{cases}
\mathbf{V}_m^{(n)}, &\hspace*{-.55cm} \text{if } \mathrm{Tr}(\mathbf{V}_m^{(n)}\big(\mathbf{V}_m^{(n)}\big)^H) \leq P_{\max}, \\[6pt]
\dfrac{\mathbf{V}_m^{(n)}}{\|\mathbf{V}_m^{(n)}\|} \sqrt{P_{\max}}, & \text{otherwise}.
\end{cases}\nonumber\\[-.6cm]\label{eq:16}
\end{align}
This approach replaces the complex operations needed for the estimation of $\mathbf{V}$ with simple operations that can be implemented as neural network layers.

\subsection{Uncertainty Injection}
In order to guarantee the solution robustness, the proposed UI-DUFP framework integrates an uncertainty injection step, a mechanism inspired by \cite{cui2023uncertainty}. The key idea is to expose the network to multiple realizations of uncertain parameters, such as channel coefficients, during training, allowing it to learn solutions that remain effective across variations.

Specifically, during each training iteration, $B$ random realizations of the channel estimation error are drawn, and the perturbed estimated channel vectors are computed as
\begin{align}
    \mathbf{\hat{h}}_{k,b} = \mathbf{h}_{k} + \mathbf{h}_{k,b}^\text{err}, &&\forall k\in\mathcal{K}, b=1,\dots,B.
\end{align}
The corresponding estimated channel matrix for sample $b$ is denoted by $\mathbf{\hat{H}}_b = [\mathbf{\hat{h}}_{1,b}, \mathbf{\hat{h}}_{2,b}, \dots, \mathbf{\hat{h}}_{K,b}]$. These sampled channel realizations are injected into the network after the beamforming update, producing $B$ corresponding objective values $\{R_{Q,b}(\mathbf{V}, \mathbf{g}, \mathbf{u}; \mathbf{\hat{H}}_b)\}_{b=1}^{B}$. The set of objective values is then sorted, and the robust objective is approximated by selecting the $\gamma \cdot B$-th element, corresponding to the empirical $\gamma$-th quantile of the WSR distribution.
The gradient of this quantile-based objective is propagated backward to update the network parameters, encouraging the unfolded model to generate beamforming solutions that maintain high performance even under adverse or uncertain channel realizations.

\subsection{Proposed Algorithm}
\begin{figure}[t]
    \centering
    \includegraphics[width=1\linewidth]{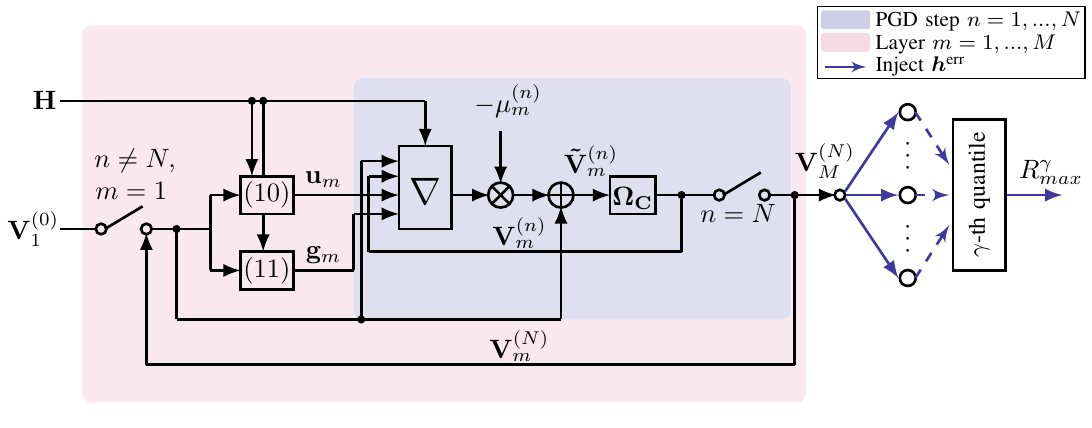}\vspace*{-.2cm}
    \caption{Flow diagram of the proposed UI-DUFP framework illustrating the input parameters (channel matrix and initial beamforming matrix), the optimized beamforming output, and the quantile-based robust WSR.}
    \label{fig:blockschaltbild}
\vspace*{-.4cm}
\end{figure} 
By integrating the DUFP structure with the uncertainty-injection scheme, we propose the UI-DUFP framework to solve the robust optimization problem in \eqref{eq:9}. Fig.~\ref{fig:blockschaltbild} depicts the overall architecture of the proposed approach, and the corresponding algorithmic steps are listed in Algorithm~\ref{alg:two}.

More specifically, the algorithm is initialized with a matched-filter beamformer $\mathbf{V}_1^{(0)} = \alpha\mathbf{H}$, where $\alpha \in \mathbb{R}$ controls the initial transmit-power scaling. For each layer $m=1,\dots,M$, the auxiliary variables $\mathbf{g}_m$ and $\mathbf{u}_m$ are first updated according to \eqref{eq:10} and \eqref{eq:11}. The beamforming matrix $\mathbf{V}_m^{(n)}$ is then refined over $N$ PGD steps as given in \eqref{eq:15}, with the transmit power constraint enforced at each step via \eqref{eq:16}. The output $\mathbf{V}_m^{(N)}$ of layer $m$ then serves as the input initialization $\mathbf{V}_{m+1}^{(0)}$ for the next layer. This process is repeated for all $M$ layers, resulting in the final beamforming matrix $\mathbf{V}_M^{(N)}$.\looseness-1

To incorporate robustness, $B$ independent realizations of the channel estimation error are generated and injected into the estimated channels, yielding the estimated channel matrices $\{\mathbf{\hat{H}}_b\}_{b=1}^{B}$. For each realization, the network computes the corresponding WSR value $\{R_{Q,b}(\mathbf{V}, \mathbf{g}, \mathbf{u}; \mathbf{\hat{H}}_b)\}_{b=1}^{B}$. The empirical $\gamma$-th quantile over these $B$ values defines the robust objective $R_{\max}^\gamma$. 
\begin{algorithm}[b]
\caption{Proposed UI-DUFP Algorithm}\label{alg:two}
\begin{algorithmic}[1]
\STATE Initialize $\mathbf{V}_1^{(0)} = \alpha\mathbf{H}$
\FOR {$m = 1,\dots,M$}
 \STATE Update $\mathbf{u}_m$ via \eqref{eq:10}
 \STATE Update $\mathbf{g}_m$ via \eqref{eq:11}
 \FOR{ {$n = 1,\dots,N$}}
  \STATE Update $\mathbf{V}_m^{(n)}$ using \eqref{eq:15}
 \ENDFOR
\ENDFOR 
\STATE Set $\mathbf{u}\hspace*{-.03cm}=\hspace*{-.03cm}\mathbf{u}_M$, $\mathbf{g}\hspace*{-.03cm}= \hspace*{-.03cm}\mathbf{g}_M$, and $\mathbf{V}\hspace*{-.03cm}=\hspace*{-.03cm}\mathbf{V}_M^{(N)}$
\STATE Sample $B$ channel errors to obtain $\mathbf{\hat{H}}_b$, $b=1,\dots,B$
\FOR{$b = 1,\dots,B$}
   \STATE Compute $R_{Q,b}(\mathbf{V}, \mathbf{g}, \mathbf{u}; \mathbf{\hat{H}}_b)$ using \eqref{eq:6}
\ENDFOR
\STATE Output robust WSR $R_{\max}^\gamma$
\end{algorithmic}
\end{algorithm}

The standard loss function for training the unfolded network is defined as the negative WSR:
\begin{equation}\label{eq:18}
\mathcal{L}{(\boldsymbol{\mu})} = - \sideset{}{_{m=1}^{M}}\sum \sideset{}{_{k\in\mathcal{K}}}\sum w_k R_Q\big(\mathbf{V}_m^{(N)}, \mathbf{g}_m, \mathbf{u}_m\big),
\end{equation}
where the beamforming matrix $\mathbf{V}_m^{(N)}$ at each layer is parameterized by the learnable step size vector $\boldsymbol{\mu} = [\mu_1^{(1)},\dots,\mu_M^{(N)}]^T$, as defined in \eqref{eq:15}. The parameters $\boldsymbol{\mu}$ are optimized during training via backpropagation to maximize the overall WSR across all unfolded layers. While the standard loss in \eqref{eq:18} maximizes the WSR across the unfolded layers, it does not enforce the probabilistic robustness constraint in \eqref{eq:8}.

To enhance robustness against channel uncertainties and mitigate potential training issues such as vanishing gradients or instability caused by extreme channel realizations \cite{pellaco2021matrix}, the loss function is extended to incorporate the $\gamma$-quantile of the WSR over multiple channel estimation error realizations. Furthermore, to stabilize gradient propagation and promote consistent performance across all layers, the WSR from each intermediate unfolding iteration is included in the overall loss. The resulting robust quantile-based loss is defined as
\begin{align}\label{eq:19}
\mathcal{L}(\boldsymbol{\mu}) 
   = & - \sum_{m=1}^{M} 
      \text{quantile}_{\gamma}\Big( 
       R_{Q}(\mathbf{V}_m^{(N)}, \mathbf{g}_m, \mathbf{u}_m;\hat{\mathbf{H}}_{1}), \notag \\ 
    &\hspace*{1.2cm} \dots, R_{Q}(\mathbf{V}_m^{(N)}, \mathbf{g}_m, \mathbf{u}_m;\hat{\mathbf{H}}_{B}) 
      \Big),
\end{align}
where $\text{quantile}_{\gamma}(\cdot)$ denotes the empirical $\gamma$-th quantile operator computed across the $B$ injected channel realizations. During backpropagation, gradients of this robust loss are propagated through all unfolded layers, allowing the network to learn parameters that yield reliable beamforming performance even under severe channel uncertainties. The quantile-based loss in \eqref{eq:19} implicitly satisfies the robustness constraint in \eqref{eq:9}. By maximizing the $\gamma$-th quantile of the WSR across multiple channel estimation error realizations, the network minimizes the probability that the achievable rate falls below $R_{\max}^{\gamma}$, thereby ensuring that the WSR satisfies the threshold at least in $(1-\gamma)$ realizations under channel uncertainty. 

\section{Simulation Results and Discussion}
In this section, we numerically evaluate the performance of the proposed UI-DUFP algorithm. The simulated network consists of a single BS equipped with $L=4$ antennas serving $K=4$ single-antenna users. The noise variance $\sigma^2$ and the user weights $w_k$ are set to $1$ for all users, and the total BS transmit power budget is fixed at $40$ dBm. The channel vectors are modeled as independent small-scale Rayleigh fading, i.e., $\mathbf{h}_k \sim \mathcal{CN}(\mathbf{0}, \mathbf{I}_L)$. We consider up to $M = 1,...,6$ unfolded layers, each employing either $N=4$ or $N=8$ PGD steps, denoted as 4PGD and 8PGD, respectively. The trainable parameters are the PGD step sizes $\mu_m^{(n)}$, initialized as $\mu_m^{(n)} = 1$, for all $m,n$. The PGD steps are updated using the Adam optimizer with a learning rate of $10^{-3}$. The training dataset comprises $8000$ batches, each containing $64$ random channel realizations, while testing is conducted over $50$ batches of equal size. For robustness evaluation, $B=1000$ realizations of the channel estimation error are injected per channel sample, with the error variance $\sigma^2_h$ varied across $[0.01, 0.05, 0.09, 0.13, 0.17]$. The quantile parameter is set to $\gamma=0.05$, corresponding to the $5$-th percentile of the WSR distribution. As performance benchmarks, two widely adopted iterative optimization algorithms, WMMSE \cite{pellaco2021matrix} and FP \cite{shen2018fractional}, are used as baselines. The WMMSE implementation is realized in PyTorch, while FP is executed in MATLAB. Additionally, we include the DL-based uncertainty injection (DL-UI) for robust sum rate maximization from \cite{cui2023uncertainty} as an additional baseline, employing a regularized zero-forcing beamformer with parameter $\alpha=1$. Building on this setup, the proposed UI-DUFP framework is evaluated in terms of its ability to improve robustness and WSR performance under uncertain channel conditions.

\subsection{WSR Performance}
\begin{figure}[t]
\centering
\includegraphics[width=1\linewidth]{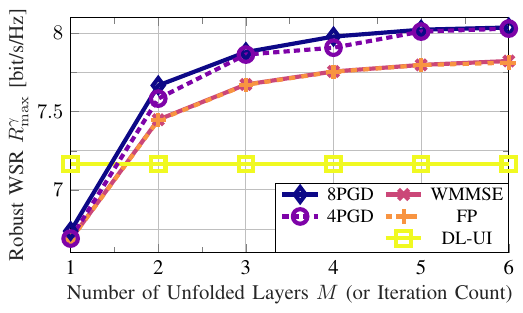}\vspace*{-.2cm}
\caption{Robust WSR vs. number of unfolded layers $M$ (UI-DUFP) and iteration count (WMMSE and FP) with $\sigma^2_h = 0.05$.}
\label{fig2}
\vspace*{-.4cm}
\end{figure}
Fig.~\ref{fig2} depicts the robust WSR $R_{\max}^\gamma$ of all evaluated schemes under channel uncertainty with variance $\sigma^2_h = 0.05$. The proposed UI-DUFP framework is trained for the 4PGD and 8PGD configurations using uncertainty injection and the $5$-th percentile ($\gamma=0.05$) loss formulation. In contrast, WMMSE and FP are optimization-based algorithms that do not involve a training phase. Instead, they compute their respective optimal beamforming solutions for each channel realization, and their performance is subsequently evaluated under the same channel uncertainty and quantile criterion to ensure a fair comparison.

As shown in Fig.~\ref{fig2}, the proposed algorithm consistently achieves higher robust WSR than the classical WMMSE and FP methods across all considered layers and iterations. In fact, each unfolded layer of the proposed network effectively mimics one iteration of the WMMSE or FP optimization, transforming the iterative process into a trainable architecture.
At the first iteration, all algorithms except DL-UI exhibit nearly identical WSR, indicating similar initial performance. As the number of iterations increases, however, the 4PGD and 8PGD variants of UI-DUFP show a clear performance gain over WMMSE and FP. Although WMMSE and FP exhibit stable and gradual WSR improvement across iterations, their values remain consistently below those of the proposed robust framework. The non-iterative baseline DL-UI is independent of the iteration count. In fact, DL-UI achieves the highest robust WSR after the first iteration but is subsequently surpassed by the proposed and other iterative schemes. This is because DL-UI optimizes only user power allocation while keeping the beamforming directions fixed, which limits its adaptability to changing channel conditions.

The results of Fig.~\ref{fig3} particularly demonstrate that training with uncertainty injection enables UI-DUFP to adapt its beamforming strategy to channel variations, achieving both higher and more robust WSR. Furthermore, 8PGD slightly outperforms 4PGD, indicating that additional gradient refinement steps enhance the optimization capability under uncertainty.\looseness-1

\subsection{Impact of Channel Estimation Error}
\begin{figure}[t]
\centering
\includegraphics[width=1\linewidth]{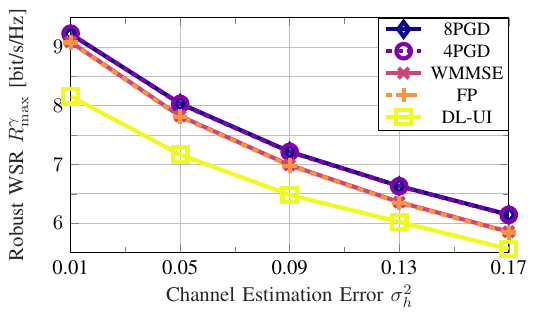}\vspace*{-.2cm}
\caption{Robust WSR as a function of $\sigma^2_h$, comparing UI-DUFP (4PGD and 8PGD) with WMMSE, FP, and DL-UI baselines.}
\label{fig3}
\vspace*{-.4cm}
\end{figure}
Fig.~\ref{fig3} illustrates the average robust WSR $R_{\max}^\gamma$ under varying channel estimation error variances $\sigma^2_h\in[0.01, 0.05, 0.09, 0.13, 0.17]$. All schemes experience a performance degradation as the channel uncertainty increases. At $\sigma^2_h = 0.01$, UI-DUFP (4PGD and 8PGD) achieves performance comparable to WMMSE and FP, with WSR values
around $8.7$ indicating similar behavior under nearly perfect channel state information. As $\sigma^2_h$ increases, however, the gap widens. For instance, at $\sigma^2_h=0.17$, WMMSE and FP achievable WSR's drop to approximately $5.74$, while UI-DUFP maintains values close to $6.0$. This demonstrates the superior robustness of the proposed approach, which preserves higher average WSR across the entire error range. The widening gap between UI-DUFP and the optimization-based benchmarks highlights the proposed algorithm's strong adaptability to uncertain channels. Fig.~\ref{fig4} further shows that, DL-UI exhibits the lowest WSR for small $\sigma^2_h$, with a relatively improved performance at higher uncertainty levels, emphasizing the benefits of incorporating uncertainty injection during training. Comparing 4PGD and 8PGD shows that additional PGD steps yield a consistent, though modest, gain. Across most error variances, 8PGD outperforms 4PGD by approximately $0.1$–$0.2$ in WSR, with a larger margin at higher uncertainty. Overall, the proposed UI-DUFP demonstrates enhanced resilience and robustness compared to conventional iterative methods under adverse channel conditions.

\subsection{Inference Time}
\begin{figure}[t]
\vspace{0.25cm}
\centering
\includegraphics[width=1\linewidth]{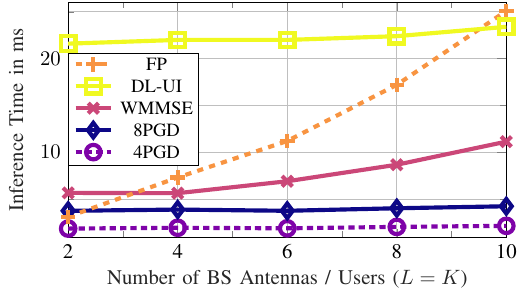}\vspace*{-.2cm}
\caption{Inference time versus number of antennas and users ($L=K$) for all evaluated schemes.}
\label{fig4}
\vspace*{-.5cm}
\end{figure}
To assess the algorithmic computational, Fig.~\ref{fig4} shows the inference time required to generate beamforming outputs (power allocations for DL-UI) for each algorithm under varying numbers of BS antennas and users ($L=K$). For averaging purposes, the inference procedure is repeated six times per configuration, and all measurements are conducted on a single CPU to ensure a fair comparison without hardware-specific acceleration. As shown in Fig.~\ref{fig4}, the proposed UI-DUFP achieves the lowest inference time among all schemes, remaining almost constant across different antenna and user settings. The 8PGD variant requires slightly more time than 4PGD due to its additional gradient steps but still scales efficiently with system size, since each PGD step corresponds to a lightweight forward pass within the network. Among the baseline methods, WMMSE exhibits the next-best runtime, increasing moderately with $L$ and $K$ because of its iterative updates. FP demonstrates the highest inference time, growing rapidly with system size due to repeated parameter updates and matrix inversions. The purely DL-based DL-UI shows relatively high inference time for small systems ($L=K\leq8$) but surpasses FP for larger configurations. Overall, DL-UI, 4PGD, and 8PGD display superior scalability compared to the optimization-based baselines, which highlights the numerical prospects of the proposed deep unfolding with uncertainty-aware algorithm, which achieves both robust performance and high runtime efficiency.\looseness-1

\section{Conclusion}
This paper proposes a UI-DUFP framework for robust beamforming in multi-user wireless networks, aiming to maximize the network WSR. Building on the FP algorithm, the proposed method first unfolds FP iterations into neural network layers, which are further refined through PGD steps. The training process is then augmented by an uncertainty injection scheme that directly introduces channel estimation errors during training to implicitly account for robustness. Simulation results demonstrate that the proposed framework achieves higher WSR performance and improved robustness to channel estimation errors compared to conventional optimization algorithms such as WMMSE and FP. The paper particularly shows that training with uncertainty injection and a quantile-based loss enables the network to adapt effectively to varying channel conditions and maintain stable performance under severe uncertainty. Increasing the number of PGD steps is also shown to enhance robustness, while inference time evaluations further highlight computational efficiency and scalability of the proposed approach. Overall, the proposed UI-DUFP algorithm proves to offer an effective and time-efficient solution for robust wireless optimization. The proposed framework, therefore, promises to be among the strong enablers of future 6G networks, as its numerical prospects reflect its strong capability at boosting both the robust performance and the system computational efficiency.\looseness-1

\bibliographystyle{IEEEtran}
\bibliography{bibliography}

\vspace{12pt}

\end{document}